\documentclass[twocolumn,prl,showpacs,preprintnumbers,amsmath,amssymb,nosuperscriptaddress,nofootinbib]{revtex4}

\usepackage{graphicx}
\usepackage{dcolumn}
\usepackage{bm}
\usepackage{color}
\usepackage{setspace}
\usepackage{amssymb}
\usepackage{tabularx}
\usepackage{hyperref}

\newcommand \A {\alpha}

\newcommand \bvec{\left( \begin{array}{c} }
\newcommand \evec{\end{array} \right)}

\newcommand \bea{\begin{eqnarray} }
\newcommand \eea{\end{eqnarray} } 
\newcommand \nn {\nonumber}
\newcommand {\be} {\begin{equation}}
\newcommand {\ee} {\end{equation}}

\newcommand {\mbx} {\mbox{}}

\newcommand {\ata} {& \times &}

\begin{document}

\title {Mass depletion: a new parameter for quantitative jet modification}

\author{A.~Majumder}\affiliation{Department of Physics and Astronomy, Wayne State University, Detroit, Michigan 48201}
\author{J.~Putschke}\affiliation{Department of Physics and Astronomy, Wayne State University, Detroit, Michigan 48201}

\begin{abstract}
We propose an extension to classify jet modification in heavy-ion collisions by including the jet mass along with its energy. The mass of a jet, as measured by jet reconstruction algorithms, is constrained by the jet's virtuality, which in turn has a considerable effect on such observables as the fragmentation function and jet shape observables. The leading parton, propagating through a dense medium, experiences substantial virtuality (or mass) depletion along with energy loss. Meaningful comparisons between surviving jets and jets produced in $p$-$p$ collisions require mass depletion to be taken into account. Using a vacuum event generator, we show the close relationship between the actual jet mass and that after applying a jet reconstruction algorithm. Using an in-medium event generator, we demonstrate the clear difference between the mass of a surviving parton exiting a dense medium and a parton with a similar energy formed in a hard scattering event. Effects of this difference on jet observables are discussed.
\end{abstract}

		
\maketitle

With the advent of the LHC, the study of the modification of hard jets in a Quark-Gluon-Plasma (QGP) has entered a detailed phase: 
Unlike the case at the Relativistic Heavy-Ion Collider (RHIC), where one has so far been limited to few particle observables~\cite{star_white,phenix_white}, 
isolation and reconstruction of particles within a jet, from the background of particles at the LHC, 
has led to an entirely new methodology of studying jets~\cite{atlasjetshape,atlasjet,cmsjet,Roland:2011cs,Chatrchyan:2011sx,rosi,marta,alicejetqm}: 
currently, new observables such as the dijet asymmetry, intra-jet fragmentation functions and jet shapes are being measured in multiple experiments. 

By all accounts, 
detailed comparisons between theory and the new observables should lead to deeper insight into the mechanisms by which jet showers are modified by the presence of a medium~\cite{Idilbi:2008vm,Majumder:2007hx,Majumder:2008zg,Majumder:2009ge,Majumder:2009zu}, 
and how energy flows away from the jet into the medium~\cite{Qin:2009uh,Renk:2013pua}. 
However, in order to make such comparisons, any calculation of jet modification has to be incorporated within an event generator\footnote{All simulations in this Letter will be carried out using the MATTER event generator~\cite{Majumder:2013re}, which is based on a medium modification~\cite{Guo:2000nz} of 
the Dokshitzer-Gribov-Lipatov-Altarelli-Parisi (DGLAP) shower~\cite{Dokshitzer:1977sg,Gribov:1972rt,Altarelli:1977zs}, or using the PYTHIA/JETSET event generators which are also based on a DGLAP shower.}. Generated events, both with and without a medium, will have to be reconstructed similarly to those in the experiment, prior to any detailed comparison.

In such comparisons, there is a marked difference between the case of single particles and full jets: 
Single particles at high-$p_{T}$ approximate to nearly massless four-vectors, 
and thus may be classified by only three intrinsic parameters: 
The $p_{T}$, the azimuthal angle $\phi$ and the rapidity $\eta$. 
For $\phi$ integrated quantities (such as $R_{AA}$), in limited ranges of rapidity, 
the $p_{T}$ represents the sole intrinsic parameter used to quantify the yield of a hard leading particle in a jet. 
High energy jets, which are reconstructed from several such vectors have, in addition, a reconstructed mass $M$, 
which is non-negligible compared to their $p_{T}$. 
To clarify our statement, if $n$ almost massless four-momenta are reconstructed, then, for the reconstructed jet, 
\bea
\vec{p}_{T} &=& \sum_{i=1}^{n} { \vec{p}_{T_{i}} } , \,\,\,\,  p_{T} =  \left|  \vec{p_{T}} \right|, \,\,\,\,  p_{T_{i}} =  \left|  \vec{p_{T_{i}}} \right| , \label{pt}
\eea
while, the $z$-component and magnitude of the vector may be obtained as,
\bea
p_{z} &=& \sum_{i=1}^{n} { p_{T_{i}} } \sinh{\eta_{i}} , \,\,\,\, 
p = \sum_{i=1}^{n} { p_{T_{i}} } \cosh{\eta_{i}}. \label{pz-p}
\eea
In terms of the equations above, the reconstructed mass of the jet may be obtained as,
\bea
M &=& \sqrt{ p^{2} - p_{T}^{2} - p_{z}^{2} } .  \label{recon-mass}
\eea
One should note that the calculation of the mass of the reconstructed jet involves, not just a knowledge of the $p_{T}$ of each particle, 
but also their azimuthal angles and their rapidities. 
As a result, full jet analyses which require a knowledge of the mass of the reconstructed jet, 
will have to be carried out at the four-vector level, as opposed to solely the $p_{T}$ level. 
This Letter represents the first in a series of attempts to generalize the phenomenology and analysis of jet modification to the four-vector level.

\begin{figure} [htb]
\vskip -0.35cm
\begin{center}
\includegraphics[width=.45\textwidth] {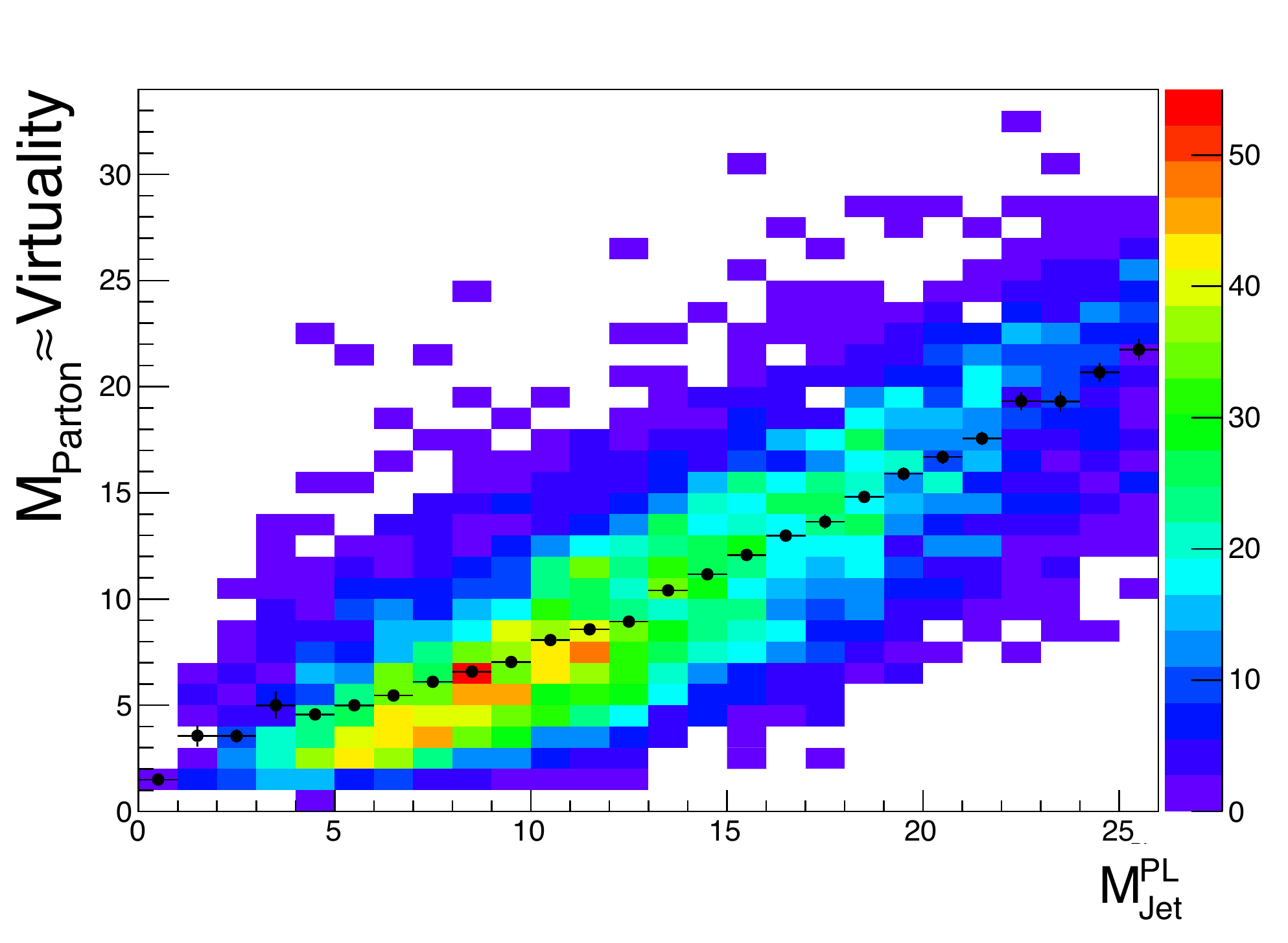}
\end{center}
\vskip -0.45cm
\caption{\label{fig:mvsm} The correlation between an underestimate of the jet mass [Eq.~\eqref{jet-mass-estimate}] and the reconstructed mass from FASTJET. Jets with energies between 90-110GeV, generated in vacuum using the PYTHIA generator. The black points represent the mean $M_{parton}$ in a given $M_{Jet}^{PL}$ bin. See text for details.}
\vskip -0.25cm
\end{figure}

In this Letter, we will use the terms virtuality, mass, reconstructed mass and scale of the jet, several times. 
The nomenclature is resolved as follows:
A hard parton, produced in a hard interaction, is almost never produced on its mass shell. 
Instead, it possesses an off-shell mass which lies between zero and $Q^{2}$, 
the off-shellness (squared) of the exchanged virtual parton that leads to the production of two back-to-back partons. 
A jet produced in Deep-Inelastic Scattering (DIS), has an off-shellness that lies between zero and $Q^{2}$, the off-shellness or virtuality of the exchanged photon. 
This value of $Q^{2}$ will be referred to as the scale of the jet, the upper limit of the virtuality of either parton that emanates from the hard collision. 
The virtuality is the property of a single parton and varies with the choice of parton in the shower. 
For a shower in vacuum, the virtuality drops monotonically with each emission. 
For a shower in a medium, the virtuality of individual partons may also rise due to scatterings. 
The mass of a jet, on the other hand, refers to the virtuality of the originating parton of the shower. 
The reconstructed mass refers to the mass obtained by applying Eqs.~(\ref{pt}-\ref{recon-mass}) on the final outgoing remnants of a shower. 
If the final remnants were massless partons, then the reconstructed mass (for a large enough capture angle $R$) would be the same as the jet mass. 
In reality, perturbative splitting is expected to continue down to a low value of virtuality $\sim 1$GeV. 
Beyond this, one expects non-perturbative processes to dominate in the hadronization of the shower. 
As such the reconstructed mass is not exactly equal to the jet mass in every event. 
However, for high enough jet energy and jet mass, the event averaged mass and reconstructed mass are closely correlated. 
This is illustrated in Fig.~\ref{fig:mvsm}, which includes the results of jets reconstructed in PYTHIA~\cite{Bengtsson:1987kr} simulations. 
The plot shows the result for the reconstructed mass versus an underestimate for the mass of the originating parton, for jets with energies between $90$-$110$GeV, 
reconstructed using FASTJET~\cite{Cacciari:2011ma} with an $R=0.7$. The mass $m$ of the parent patron with momentum $p$ 
is estimated using,
\bea
m^{2} = \left( \left| \vec{p}_{1} \right|+ \left| \vec{p}_{2}\right| \right)^{2} - \left| \vec{p}_{1} + \vec{p}_{2} \right|^{2} , \label{jet-mass-estimate}
\eea 
where $\vec{p}_{1},\vec{p}_{2}$ are the 3-momenta of the daughter partons.
This equation assumes that both daughter partons are massless and is thus an underestimate for the jet mass.

\begin{figure} [htb]
\includegraphics[width=.45\textwidth]{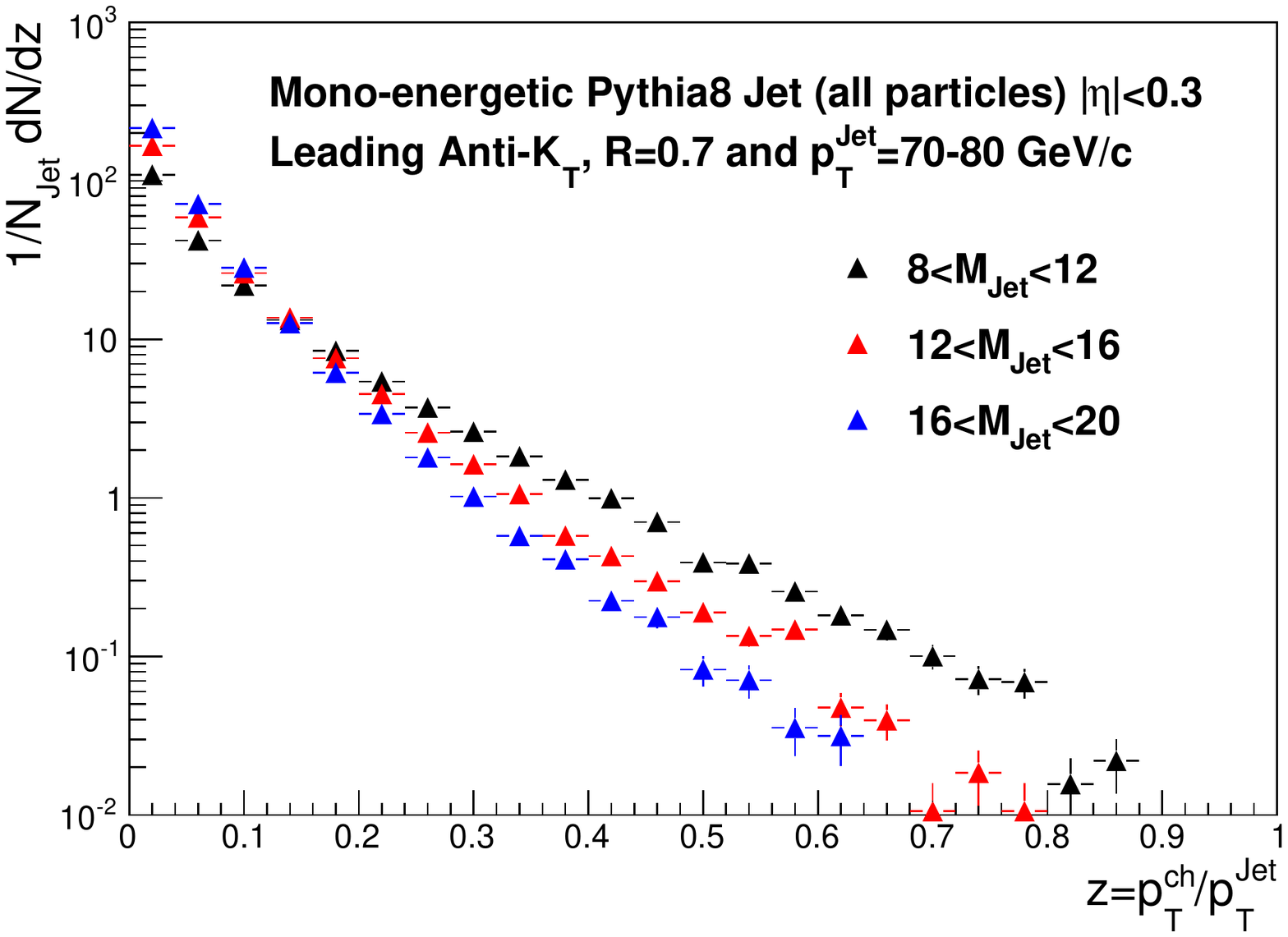}
\caption{\label{fig:scale} Pythia8 fragmentation function: for $R=0.7$ jet with $p_T^{Jet}=70$-80GeV/c for different jet mass ($M_{Jet}$) selections.}
\vskip -0.25cm
\end{figure}

Using the reconstructed mass, one may easily demonstrate the influence of jet mass on the intra-jet quantities such as the fragmentation function. 
Currently, there are multiple measurements of this quantity from LHC experiments. 
We present results for the intra-jet fragmentation function for jets with energies between 70-80 GeV, simulated by the PYTHIA event generator. 
These are reconstructed using FASTJET, in separate bins of reconstructed mass: 8-12GeV, 12-16GeV, 16-20GeV. Results are displayed in Fig.~\ref{fig:scale}. 
One will immediately note that the fragmentation function is very sensitive to the range of reconstructed mass, with larger masses leading to a suppression of 
the higher $z$ portion of the fragmentation function, and an enhancement of the lower $z$ portion. As such, the effect of reconstructed mass on the fragmentation function 
is qualitatively similar to that of the scale of the fragmentation function.

As the hard jet propagates through vacuum, the virtuality of partons tends to drop with increasing time as more partons are formed. 
In the limit that the virtuality is large compared to $\Lambda_{QCD}$, multiple radiations tend to be virtuality ordered 
and can be simulated in event generators by sampling a perturbatively calculable Sudakov form factor. 
In a medium, scattering processes may tend to increase the virtuality temporarily. 
However, on the whole, it tends to drop. 
In the limit that the virtuality is large, the jet resolves very short distance scales in the medium. 
At this resolution, the medium will appear to be diluted and thus the jet will not 
scatter as much with the constituents of the medium. 
In this limit, the few scattering Higher-Twist expressions for single gluon emission~\cite{Majumder:2009ge,Majumder:2009zu}  
are valid and one obtains virtuality ordered emissions. 
Given these approximations, 
the medium modified Sudakov form factor for a quark with light-cone momentum $p^{-}$, 
with a maximum virtuality $Q^{2}$, propagating through a dense medium, 
starting at the light-cone location $\zeta_{i}^{-} $ is given as~\cite{Majumder:2013re}, 
\bea
\mbx\!\!\!\!&& S_{\zeta_{i}^{-}}(Q_{0}^{2},Q^{2}) = \exp\left[ - \int\limits_{2 Q_{0}^{2}}^{Q^{2}} \frac{d\mu^{2}}{\mu^{2}} \frac{\A_{S}(\mu^{2})}{2 \pi}  \right.  
\label{in-med-sud} \\
 \mbx\!\!\!\! \ata \left. \int\limits_{Q_{0}/Q}^{1-Q_{0}/Q} \!\!\!\!\!dy  P_{qg}(y)  
 \left\{  1 + \int_{\zeta_{i}^{-}}^{\zeta_{i}^{-} + \tau^{-}} \!\!\!\!\!d \zeta K_{p^-,Q^2} ( y,\zeta) \right\} \right] . \nn  
\eea

In the equation above, the factor $P_{qg}(y)$ represents the splitting function for the quark to radiate a gluon with a 
momentum fraction $y$. 
The factor of unity within the curly brackets on the second line represents the contribution from 
vacuum like splits, i.e., splits which are not induced by the medium, or interfere with splits induced by the medium. 
The second term, which involves a position integral, is the contribution from the interference between vacuum like 
radiation and that induced by the medium. 
The leading twist contribution to the multiple scattering, single emission kernel $K$, is given as 
\bea
K_{p^-,Q^2} ( y,\zeta) = \frac{2 \hat{q} }{Q^{2}}
\left[ 2 - 2 \cos\left\{ \frac{Q^2 (\zeta - \zeta_i)}{ 2 p^- y (1- y)} \right\}  \right] . \label{kernel}
\eea
In the equation above, $\hat{q}$ represents the one transport coefficient of the medium, which both induces radiation and 
leads to transverse momentum diffusion of radiated gluons. Using this set up we may study the effect of the medium on the surviving energy 
and mass of hard jets. 

\begin{figure} [htb]
\includegraphics[width=.4\textwidth]{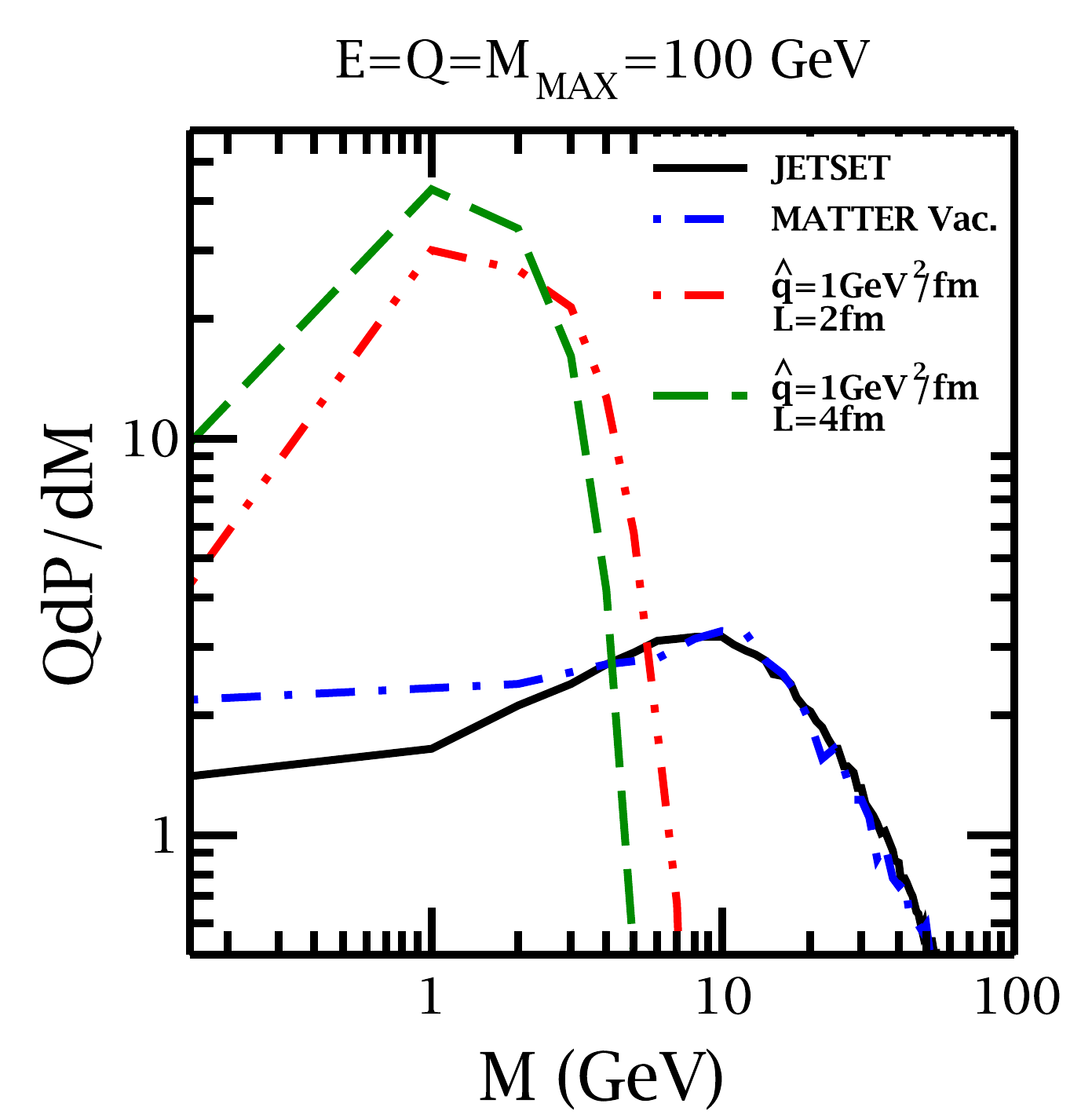}
\caption{\label{fig:LHCvirt} Virtuality distribution of the leading parton at $L=0$fm (parton that originates the shower), from MATTER, compared with the results from JETSET. Virtuality distribution of the leading parton at L=2fm (red dot-dot-dashed line) and 4fm (green dashed line). The leading parton is seen to drop virtuality swiftly with distance. 
} 
\vskip -0.25cm
\end{figure}

Along with the minima of the $y$ integration, we also insist that the first split produces two partons that are emitted in the same hemisphere.
In what follows, we present results for the theoretical process of a hard jet produced at the edge of a cubic container containing a static QGP with a fixed $\hat{q}$. 
To mimic LHC like conditions we will use a $\hat{q}=1$GeV$^{2}$/fm. This was 
found to be the approximate average value in a recent phenomenological comparison~\cite{Burke:2013yra} for 0-5\% central LHC collisions.  
We first plot the distribution of invariant mass of the originating parton for a 100GeV jet, simulated within the MATTER event generator (blue dot-dashed line) and compare with the JETSET event generator (solid black line) in Fig.~\ref{fig:LHCvirt}. Both generators yield similar distributions with small differences at very low mass $m\lesssim 1$ GeV.  One may notice that these distributions are somewhat lower than those obtained from a PYTHIA~\cite{Bengtsson:1987kr} simulation or even those measured in $p$-$p$ collisions at the LHC~\cite{ATLAS:2012am}. This is due to that fact that both JETSET and MATTER (at this stage) simulate events with no initial state.

Based on current data from the LHC experiments, the emerging picture of jet modification indicates that hard jets tend to radiate a 
fraction of their energy to softer gluons, which are then scattered out of the jet cone by the medium. The remnant hard core of the jet emerges 
from the medium with a minimal amount of net transverse momentum. The energy lost in radiated gluons is recovered at large angles away from the 
jet cone. We simulate an extreme example of this principle: After traversing a certain length of medium, all but the leading parton is stopped by the medium. 
The leading parton escapes, showers and hadronizes as a jet in vacuum, and is then captured solely and entirety by jet reconstruction and subtraction algorithms.  
Our analysis indicates that these emerging jets not only have lost a part of their energy, but compared to jets of the same lower energy, produced 
in $p$-$p$ collisions, have a much lower mass or virtuality.

\begin{figure} [thb]
\includegraphics[width=.4\textwidth]{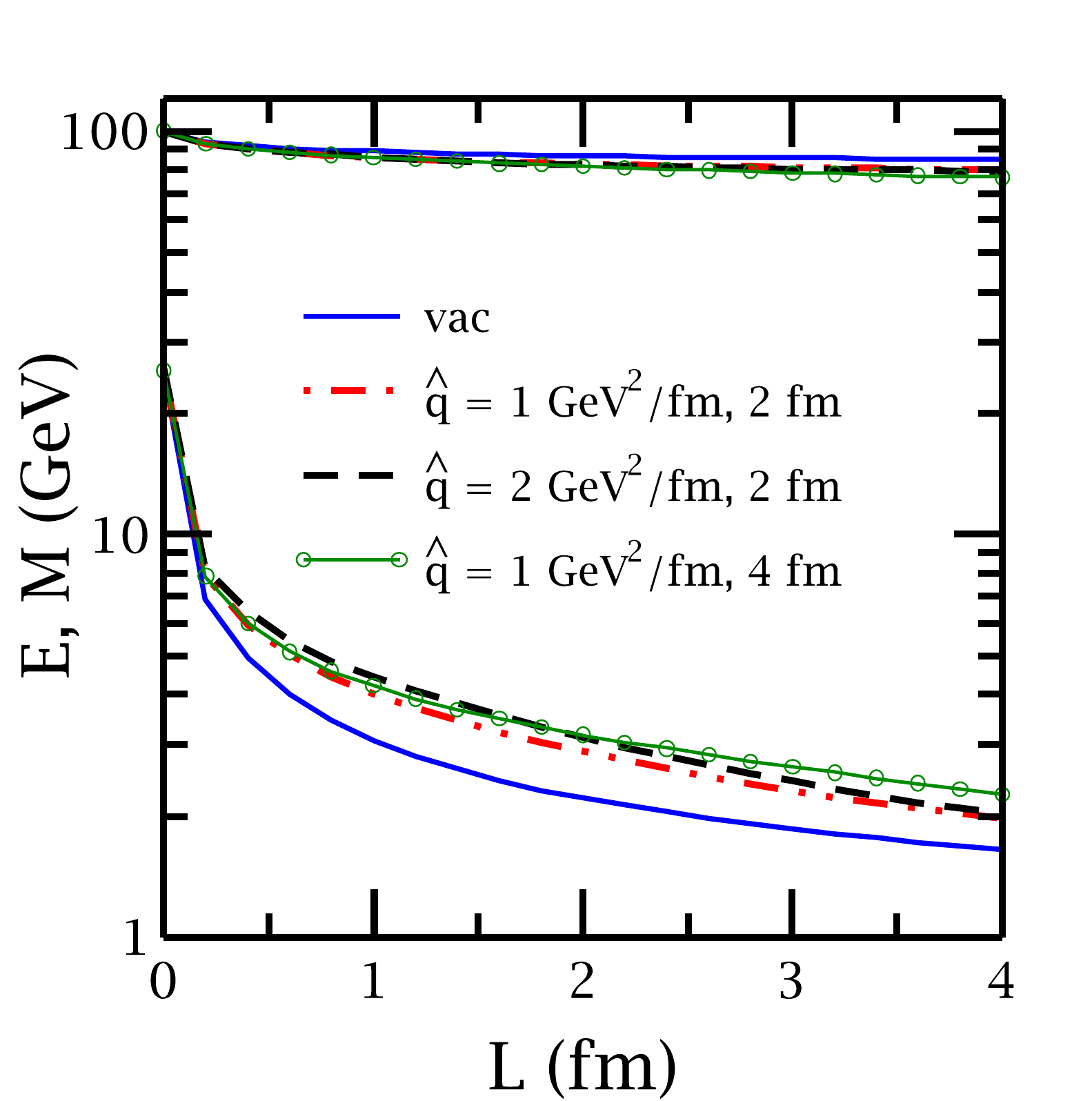}
\caption{\label{fig:LHCvirtE} The mean energy and mass of the leading parton in a quark jet as a function of length in four different scenarios. See text for details.
}
\vskip -0.25cm
\end{figure}

This is demonstrated in a MATTER simulation by allowing the produced 100 GeV jets to propagate through a static medium with a mean $\hat{q}=1$GeV$^{2}$/fm (to mimic LHC conditions).
As the shower splits into more partons, we track the leading parton. The virtuality distributions of the leading partons are then plotted in Fig.~\ref{fig:LHCvirt}, after traversing a 2fm (red dot-dot-dashed line) and 
a 4fm (green dashed line) medium. One notes that the mass of the leading parton has dropped significantly.
We contend that this large ``mass depletion'' is \emph{not} due to medium induced radiation but is mostly a DGLAP shower effect; in fact, scattering in the medium reduces the mass depletion with increasing length. To illustrate this, we plot, in Fig.~\ref{fig:LHCvirtE}, the mean mass and energy of the leading parton for four different cases: 
in vacuum and in three different media.
The lines at the top of the plot represent the energy of the leading parton, while those at the bottom of the plot represent the mass of the leading parton. 

Focussing on the mass of the leading parton, one notices that a jet in vacuum (blue solid line) loses its mass swiftly, and by a distance of 4fm, the mass has dropped below 2GeV. At this point, the leading parton still carries $\sim 90$\% of the energy of the jet on average.  
In the presence of a medium, multiple scattering tends to slow down the drop, thus the red dot-dot-dashed line ($\hat{q}=1$GeV$^{2}$/fm, $L=2$fm) starts to separate from the vacuum curve within a fraction of a Fermi. This continues to separate 
until the extent of the medium (2fm) and then begins to fall towards the vacuum curve. The case where the medium extends to $L=4$fm (green circles), the mass of the leading parton continues to separate from the vacuum curve till 4fm. For the case with double the density $\hat{q}=2$GeV$^{2}$/fm and L=2fm, the in-medium mass of the leading parton separates more swiftly from the vacuum case up to $L=2$fm and then also begins to dip towards the vacuum curve. For the case of 100GeV jets, media with realistic values of $\hat{q}$ seem to have a very minor effect on the mass depletion of the leading parton with length. 

For the case of the jet in vacuum, a mass measurement would, in principle, include most of the radiated partons and thus yield the full mass at $L=0$fm. However, in the presence of a dense medium where the radiated partons may be lost or scattered out of the jet cone, one will reconstruct a subset of the entire jet around the leading parton. In the extreme case that only the leading parton were to escape the medium, and then shower in vacuum, one will reconstruct the mass of the leading parton at the point of exit. The fragmentation of such a jet will no doubt produce a harder spectrum of hadrons compared to a jet of the \emph{same} reduced energy produced in a $p$-$p$ collision, which will start with a much larger mass.

Such a study clearly indicates the importance of jet mass measurements. 
Leaving more sophisticated treatments for future efforts, we have established that 
jets escaping a dense medium are different from those of the same energy produced in a hard collision 
in vacuum. 
Current measurements of the jet fragmentation function in $Pb$-$Pb$ compared to $p$-$p$ at the LHC~\cite{atlasFF,cmsFF} show a suppression at intermediate-$z$ and a rise towards unity (or above) at high-$z$. The intermediate-$z$ suppression is expected due to partonic energy loss. The rise at high-$z$ could be understood due to a reduced jet mass of the escaping jets in $Pb$-$Pb$ (Fig.~\ref{fig:LHCvirtE}). This would result in a harder fragmentation of the escaping jets with respect to the vacuum case at the same jet energy (Fig.~\ref{fig:scale}).  In a broader context, having access to the virtuality evolution via jet mass measurements adds a new, not yet experimentally accessed,  dimension to jet quenching measurements, by constraining both of the relevant quantities, energy and virtuality, and is expected to provide non-trivial tests for models of in-medium shower evolution. 

The authors thank, R. Reed, T. Renk, M. Verweij, and members of the JET collaboration for helpful discussions. This work was supported in part by the NSF under grant no. PHY-1207918 and the U.S. D.O.E.~(SC), Office of Nuclear Physics under award no. DE-FG02-92ER-40713.


\vskip -0.37cm
\bibliography{ref_new_used,refs_other} 

\end{document}